\documentclass[lettersize,journal]{IEEEtran}
\usepackage{amsmath,amsfonts}
\usepackage{algorithmic}
\usepackage{algorithm}
\usepackage{array}
\usepackage[caption=false,font=normalsize,labelfont=sf,textfont=sf]{subfig}
\usepackage{textcomp}
\usepackage{stfloats}
\usepackage{url}
\usepackage{verbatim}
\usepackage{graphicx}
\usepackage{cite}
\usepackage{xcolor}
\usepackage{multirow}
\usepackage{tikz}
\usetikzlibrary{arrows.meta,positioning,patterns}
\usepackage{booktabs} 

\begin{document}

\title{Adaptive Fault Injection Planning for Multi-Layer\\
Self-Healing AI Infrastructure}

\author{Saurabh Kulkarni\thanks{Saurabh Kulkarni is a Systems Integration Engineer, working at Meta Platforms, Inc., 1 Meta Way, Menlo Park, CA 94025, USA}, Yuxin Yang\thanks{Yuxin Yang is a PhD candidate in the School of Systems Science and Industrial Engineering at Binghamton University, State University of New York, Binghamton, NY 13902, USA}, Rohan Kulkarni\thanks{Rohan Kulkarni is a Engineering Manager, working at Meta Platforms, Inc., 1 Meta Way, Menlo Park, CA 94025, USA}, and Gautam Nayak\thanks{Gautam Nayak is a Manufacturing Quality Engineering Manager, working at Meta Platforms, Inc., 1 Meta Way, Menlo Park, CA 94025, USA}}


\maketitle

\begin{abstract}

Modern GPU-accelerator platforms rely on multi-layer self-healing pipelines that span hardware, firmware, management software, and orchestration. When faults propagate across layer boundaries, they can bypass detection, corrupt diagnosis, or trigger conflicting remediations—yet conventional fault-injection campaigns test each layer in isolation. We present ADA-ST, an adaptive fault-injection methodology that uses a weighted fault-propagation graph to guide cross-layer scenario selection. We construct four-layer graphs for three successive platforms at a hyperscale operator: Alpha, Beta, and Gamma. Platform Alpha, a production system that accumulated 72,550 repair tickets over four years, provides the empirical foundation; 49\% of those tickets involve cross-layer fault propagation. We show that existing static test campaigns cover only 20–25\% of the modeled fault-propagation edges, leaving approximately three-quarters of the cross-layer attack surface unexercised. ADA-ST closes this gap through iterative, activity-guided scenario selection that maximizes marginal coverage gain per iteration, reaching full edge coverage within 10 iterations on Alpha, 12 on Beta, and 9 on Gamma. The Fault-Layer Abstraction Mapping (FLAM) transfers propagation knowledge across hardware generations with 100\% fidelity from Alpha to Beta and 96\% from Beta to Gamma. Physical spot-validation on the newest platform confirms all four tested propagation edges, revealing cross-layer vulnerabilities spanning telemetry blind spots, absence-based detection gaps, multi-signal correlation failures, and trust-without-verification propagation at the L2-to-L3 boundary.
\end{abstract}


\subsection*{Notation List} \newcommand{\notentry}[2]{\par\noindent\hangindent=1.7cm\hangafter=1 \makebox[1.7cm][l]{#1}#2} \notentry{$B$}{Scenario budget} \notentry{$\mathrm{Blast}(S_k)$}{Blast radius of scenario $k$} \notentry{$C_{\text{campaign}}$}{Edge coverage ratio of a fault injection campaign} \notentry{$C_{\text{weighted}}$}{Severity-weighted edge coverage} \notentry{$d$}{Propagation latency along an edge} \notentry{$E$}{Propagation edges} \notentry{$G$}{Fault propagation graph} \notentry{$M_p$}{Mapping from roles to components for platform $p$} \notentry{$p$}{Conditional fault propagation probability} \notentry{$\mathcal{P}$}{Fault classes} \notentry{$R$}{Set of abstract functional roles in FLAM} \notentry{$S_k$}{Risk-weighted priority score for scenario $k$} \notentry{$V$}{Fault-observable vertices} \notentry{$v_{i,j}$}{Vertex at layer $i$, subsystem index $j$} \notentry{$w_1, w_2, w_3$}{Scoring weights} \notentry{$\varepsilon$}{Marginal coverage gain threshold}

\subsection*{Acronyms List} \renewcommand{\notentry}[2]{\par\noindent\hangindent=1.7cm\hangafter=1 \makebox[1.7cm][l]{#1}#2} \notentry{ADA-ST}{Adaptive Steering Algorithm} \notentry{BMC}{Baseboard Management Controller} \notentry{DVT}{Design Validation Testing} \notentry{FLAM}{Functional Layer Abstraction Model} \notentry{GPU}{Graphics Processing Unit} \notentry{HMC}{Hardware Management Controller} \notentry{MAPE-K}{Monitor, Analyze, Plan, Execute over a shared Knowledge base} \notentry{MP}{Mass Production} \notentry{NIC}{Network Interface Controller} \notentry{NPI}{New Product Introduction} \notentry{PCIe}{Peripheral Component Interconnect Express} \notentry{RMC}{Rack Management Controller} \notentry{SAAG}{Signal Aggregator} \notentry{SEU}{Scenario Effort Unit}

\section{Introduction}

\IEEEPARstart{M}{odern} AI training infrastructure operates at unprecedented scale. A single training cluster may contain tens of thousands of GPUs interconnected by high-bandwidth fabrics, cooled by facility-scale liquid-cooling systems, and managed by layered software stacks that automatically detect, diagnose, and remediate hardware failures \cite{jiang2024, kokolis2025, dong2025}. We organize these \emph{self-healing mechanisms} into four functional layers, as shown in Table \ref{tab:layers}, following the dependability taxonomy of Avizienis et al. \cite{avizienis2004}. These layers interact through well-defined interfaces: L1 hardware generates physical signals (temperature, voltage, error codes), L2 firmware samples these signals and exposes them via management protocols, L3 management plane aggregates signals into diagnostic assessments, and L4 orchestration acts on those assessments to drain, repair, or return capacity. Together, these mechanisms are important for maintaining the availability required by multi-week training jobs that cannot tolerate unplanned interruptions. Recent studies document the severity of reliability challenges in such clusters, including GPU failure rates \cite{tiwari2015, kokolis2025}, silent data corruptions \cite{dixit2021}, and cascading training interruptions \cite{jiang2024, dong2025}.

\begin{table*}[htp] 
\caption{Four-layer self-healing architecture.} 
\label{tab:layers} 
\centering 
\begin{tabular}{@{}llll@{}} \hline 
Layer & Name & Function & Examples \\ \hline 
L1 & Hardware & Generate fault signals & GPUs, PSUs, cables \\ 
L2 & Firmware & Monitor hardware and expose telemetry & BMC, GPU and NIC firmware \\ 
L3 & Management & Collect telemetry \& diagnose faults & Health checkers \\ 
L4 & Orchestration & Schedule remediation \& restore capacity & Auto-remediation, job scheduler \\ \hline 
\end{tabular} 
\end{table*}

During New Product Introduction (NPI), platforms undergo validation campaigns that exercise individual components. These campaigns include hardware stress tests that verify GPU reliability, firmware tests that check sensor polling correctness, and software tests that confirm management daemon behavior. These campaigns are \emph{static}: they are derived from component specifications, executed in a fixed order, and evaluated against predetermined pass/fail criteria. We identify three structural limitations of this static approach:

\begin{enumerate} 
\item \textbf{No coverage model:} Static campaigns lack a formal model of what they test. Without a structured representation of fault-propagation paths, we cannot measure which cross-layer interactions have been exercised, which remain unexplored, or where the highest-risk gaps exist. 
\item \textbf{Single-layer scope:} Static campaigns test each layer in isolation. A hardware stress test confirms that a GPU fails predictably under thermal stress, but does not verify that the firmware correctly reports this failure, that the management plane correctly diagnoses the root cause, or that the orchestration layer correctly remediates the affected server without disrupting neighboring workloads. 
\item \textbf{Fixed scenario catalogs:} Static test scenarios are designed upfront from specifications and expert knowledge. These scenarios cannot anticipate emergent failure modes that arise from the interaction of independently correct components. For example, the firmware may correctly report both fault A and fault B on the same server. The management plane may correctly diagnose fault A and initiate a remediation action, but this action may conflict with a concurrent remediation action that the orchestration layer independently triggers in response to fault B.
\end{enumerate}

Despite a four-year operational history and 72,550 repair tickets on an AI platform alone, we find that the existing static fault-injection campaigns exercise only around 20-25\% fault-propagation edges. However, 49\% of the field tickets involve cross-layer propagation—failures that traverse boundaries between hardware, firmware, management software, and orchestration—yet the static campaigns test exclusively within single-layer detection paths. Cross-layer failures are not merely more numerous; they are disproportionately resource intensive. Tickets spanning all four layers require an average of 36.5 days to resolve—6.6 times longer than single-layer tickets (5.5 days). This resolution-time multiplier demonstrates that cross-layer propagation creates diagnostic complexity that single-layer testing cannot anticipate or exercise. This structural coverage gap motivates an adaptive approach. We propose an adaptive, multi-layer fault-injection planning methodology that addresses all three limitations:

\begin{enumerate}
\item A \textbf{fault propagation graph} that models observability points and propagation edges across the hardware, firmware, management, and orchestration layers, providing a formal coverage model for the self-healing pipeline.
\item An \textbf{Adaptive Campaign Steering Algorithm} (ADA-ST) that uses runtime coverage feedback to select the next injection scenario, prioritizing unexplored graph edges and spawning exploratory variants when it observes unexpected cascades. ADA-ST is conceptually related to coverage-guided fuzzing \cite{bohme2016} but operates at infrastructure scale.
\item An \textbf{empirical characterization of cross-layer failure patterns} in production AI training infrastructure, drawn from the repair-ticket corpus.
\end{enumerate}

We also contribute a Functional Layer Abstraction Model (FLAM) that decouples fault scenarios from specific hardware, enabling campaign reuse across platform generations with only a mapping-table update. We evaluate the methodology using a retrospective–prospective hybrid design across three anonymized platforms. For Platform Alpha (in production), we reconstruct the static campaign's graph coverage and use production incidents to identify ground-truth blind spots. Retrospective simulation shows that ADA-ST covers all 27 blind-spot edges within 10 iterations. For Platform Beta (in NPI), we construct the fault-propagation graph from architecture documents and demonstrate 100\% scenario transfer via FLAM. The adaptive campaign covers 100\% of graph edges, compared with 24.1\% for the static baseline. For Platform Gamma (in early NPI), we analyze its 384-test NPI plan and show that it achieves 25.0\% edge coverage—within the same 20–25\% band as Alpha's 233-test plan (20.6\%). This result confirms that the structural limitation persists regardless of test-plan size or platform maturity. Overall, ADA-ST achieves complete edge coverage within 9–12 iterations per platform while static NPI campaigns plateau at 20–25\%.

The remainder of this paper is organized as follows. Section II surveys related work. Section III defines the system model, fault-propagation graph, ADA-ST algorithm, FLAM abstraction and five-leg evaluation framwork. Section IV presents the results and Section V discusses limitations. Section VI presents conclusions and future work.

\section{Related Work}
\label{sec:related}

\subsection{Infrastructure Reliability and Failure Analysis}

Large-scale AI training clusters exhibit distinctive reliability challenges due to unprecedented GPU density, high-bandwidth interconnects, and multi-week job durations. Kokolis et al.~\cite{kokolis2025} analyze 11 months of operational data from two production ML clusters, covering over 150 million GPU hours and establishing failure taxonomies and predictive models for GPU-dense environments. Hu et al.~\cite{hu2024} characterize LLM development workloads across a six-month datacenter trace, documenting frequent hardware failures and the need for fault-tolerant pretraining. Jeon et al.~\cite{jeon2019} analyze multi-tenant GPU clusters, revealing GPU memory errors and scheduling failures as dominant causes of workload interruption. Tiwari et al.~\cite{tiwari2015} report reliability lessons from GPU experience on the Titan supercomputer, quantifying failure rates at HPC scale.

At the system level, Jiang et al.~\cite{jiang2024} demonstrate the operational challenges of training LLMs at scales exceeding 10,000 GPUs in MegaScale, emphasizing fault tolerance, straggler mitigation, and deep observability. Dong et al.~\cite{dong2025} present C4, a production system for real-time anomaly detection during AI training that rapidly identifies faulty components and triggers automated recovery. Dixit et al.~\cite{dixit2021} document silent data corruptions at fleet scale, identifying hardware-induced errors that bypass traditional detection and span the L1--L2 boundary in our model. Nie et al.~\cite{nie2024} demonstrate that reliability-aware job placement in training clusters can mitigate hardware-failure impact, complementing our focus on testing the self-healing pipeline's response to such failures.

These AI-specific findings build on a broader tradition of empirical failure analysis. Schroeder and Gibson~\cite{schroeder2007} study disk failures, Schroeder et al.~\cite{schroeder2009} study DRAM errors, and Sridharan et al.~\cite{sridharan2015} study memory errors at scale, collectively establishing that real-world failure rates substantially exceed vendor specifications. Vishwanath and Nagappan~\cite{vishwanath2010} extend this methodology to cloud hardware reliability, and Ford et al.~\cite{ford2010} study availability in globally distributed storage systems, analyzing correlated and cascading failures. In the network domain, Gill et al.~\cite{gill2011} characterize failures in production datacenters, and Potharaju and Jain~\cite{potharaju2013} empirically study cloud network failures and their service impact. Oppenheimer et al.~\cite{oppenheimer2003} find that operator error and misconfiguration dominate internet service failures, motivating systematic testing of operational procedures. Gunawi et al.~\cite{gunawi2016} analyze hundreds of cloud outages and find that automated responses cause approximately 30\% of cascading failures.

These works model fault propagation within single domains (network, storage, cloud services, or AI hardware). Our graph spans four functional layers with empirically derived conditional probabilities, designed not for monitoring or diagnosis but for guiding test-campaign design. Our work addresses the complementary question: how to systematically test the self-healing response to documented failures before they manifest in production.

\subsection{Fault Injection Techniques} 

Fault injection has a rich history in dependability validation. Arlat et al.~\cite{arlat1990} establish foundational methodology for physical, software-implemented, and simulation-based injection, defining coverage and latency metrics. Arlat et al.~\cite{arlat1993} formalize the concept of a fault-injection campaign and propose metrics for evaluating campaign completeness. Natella, Cotroneo, and Madeira~\cite{natella2016} provide a comprehensive survey of software fault injection spanning methodology, tools, and applications. Vieira and Madeira~\cite{vieira2002} extend campaign design with formal coverage-completeness criteria.

Carreira et al.~\cite{carreira1998} introduce feedback-driven parameter adjustment in Xception, adapting injection timing and location to improve campaign efficiency. Their adaptation operates within a single-layer model, adjusting injection parameters rather than selecting cross-layer scenarios. At the hardware level, Hari et al.~\cite{hari2017} develop SASSIFI for architecture-level GPU fault injection, evaluating application resilience to transient hardware faults. SASSIFI operates at the single-device level and does not model cross-layer propagation. Our methodology extends the adaptive concept to multi-layer scenario selection at infrastructure scale (rack-scale, four layers), and can incorporate single-device tools as injection primitives within broader multi-layer scenarios.

Chaos engineering~\cite{basiri2016}, \cite{rosenthal2020} injects faults in production to validate resilience. Alvaro et al.~\cite{alvaro2015} propose Lineage-Driven Fault Injection (LDFI), which uses data lineage to identify minimal fault combinations that can affect outputs, providing a principled alternative to random injection. Gunawi et al.~\cite{gunawi2011} propose FATE and DESTINI for systematic cloud recovery testing. Cotroneo et al.~\cite{cotroneo2020} present ProFIPy, a programmable fault-injection framework offering as-a-service injection for cloud environments. These approaches share our philosophy of proactive fault injection but operate in the software domain and do not model cross-layer propagation spanning hardware, firmware, management, and orchestration.

Most relevant to our approach, Cotroneo et al.~\cite{cotroneo2022} propose fault-injection analytics, using data analysis to discover failure modes in cloud systems, as compared in Table~\ref{tab:comparison}. Their analytics-driven method shares our philosophy of leveraging operational data to guide injection campaigns. However, their approach operates within a single service layer and does not model multi-layer propagation graphs or provide formal coverage metrics across infrastructure layers. Our coverage-driven steering is also conceptually related to coverage-guided greybox fuzzing~\cite{bohme2016}, which models exploration as a Markov chain and demonstrates that coverage-guided approaches discover more defects than random injection. ADA-ST extends both of these data-driven philosophies with explicit graph-coverage guidance and cross-layer scenario steering. The formal convergence properties differ: the guarantee in~\cite{bohme2016} is probabilistic over a continuous input space, whereas ADA-ST operates over a finite, known graph with deterministic scenario-to-edge mappings (see Section~\ref{sec:method}).

\begin{table}[!h] 
\caption{Comparison of Cotroneo et al.~\cite{cotroneo2022} and ADA-ST.} \label{tab:comparison} 
\centering 
\begin{tabular}{@{}lll@{}} \hline 
Dimension & Cotroneo et al.~\cite{cotroneo2022} & ADA-ST \\ \hline 
Layers & Single service layer & Four infrastructure layers \\ Coverage & Fault-load representativeness & Graph-edge coverage (Eq.~\ref{eq:coverage}) \\ 
Scale & Cloud microservices & Hyperscale AI clusters \\ 
\multirow{2}{*}{Domain} & \multirow{2}{*}{Software faults} & Hardware, firmware, \\ & & software, orchestration \\ 
\multirow{2}{*}{Adaptation} & \multirow{2}{*}{Analytics-driven fault load} & Runtime coverage-guided \\ & & steering \\ 
Reuse & Not addressed & FLAM abstraction model \\ \hline 
\end{tabular} 
\end{table}

\subsection{Dependability Foundations and Self-Healing}

Avizienis et al.~\cite{avizienis2004} establish the foundational taxonomy of dependable and secure computing, defining faults, errors, and failures and their propagation relationships. Our four-layer fault-propagation graph builds on this taxonomy, instantiating fault–error–failure chains across infrastructure layers. The autonomic computing vision~\cite{kephart2003} defines the MAPE-K control loop (Monitor, Analyze, Plan, Execute) for self-healing systems. Our four-layer model is a manifestation of this architecture for AI infrastructure: L1 hardware generates events (Monitor), L2 firmware and L3 management plane analyze and diagnose (Analyze, Plan), and L4 orchestration remediates (Execute). Ghosh et al.~\cite{ghosh2007} survey self-healing architectures across domains. These works design and analyze self-healing systems, but none proposes a methodology for testing whether the self-healing pipeline works correctly under cascading failures. Our contribution fills this gap.

Cai et al.~\cite{cai2017} develop a Bayesian-network-based fault-diagnosis methodology for complex systems. While Bayesian networks provide probabilistic fault diagnosis within a single system, our graph models fault propagation across infrastructure layers for test planning rather than diagnosis. The IEC 61025 fault-tree analysis standard~\cite{iec61025} and IEC 60812 FMEA standard~\cite{iec60812} provide complementary frameworks. Our fault taxonomy is consistent with FMEA fault-effect chains, extended to capture cross-layer propagation in AI infrastructure. The scoring-function weights follow the MCDA weighting philosophy of IEC 60812 risk priority numbers, where frequency, consequence, and detectability are weighted comparably.

To our knowledge, no prior work combines (1)~a multi-layer fault propagation graph for AI infrastructure, (2)~coverage-driven adaptive campaign steering, (3)~cross platform scenario transfer via functional abstraction, and (4)~empirical validation at hyperscaler scale (72,550+ tickets, 4 years). Each element has a precursor, which are fault injection analytics~\cite{cotroneo2022} for data driven campaign design, coverage guided fuzzing~\cite{bohme2016} for feedback driven exploration, abstraction models in fault injection surveys~\cite{natella2016}. However, their combination for validating self-healing infrastructure is novel.

\section{Methodology}
\label{sec:method}

\begin{figure*}
    \centering
    \includegraphics[width=0.85\linewidth]{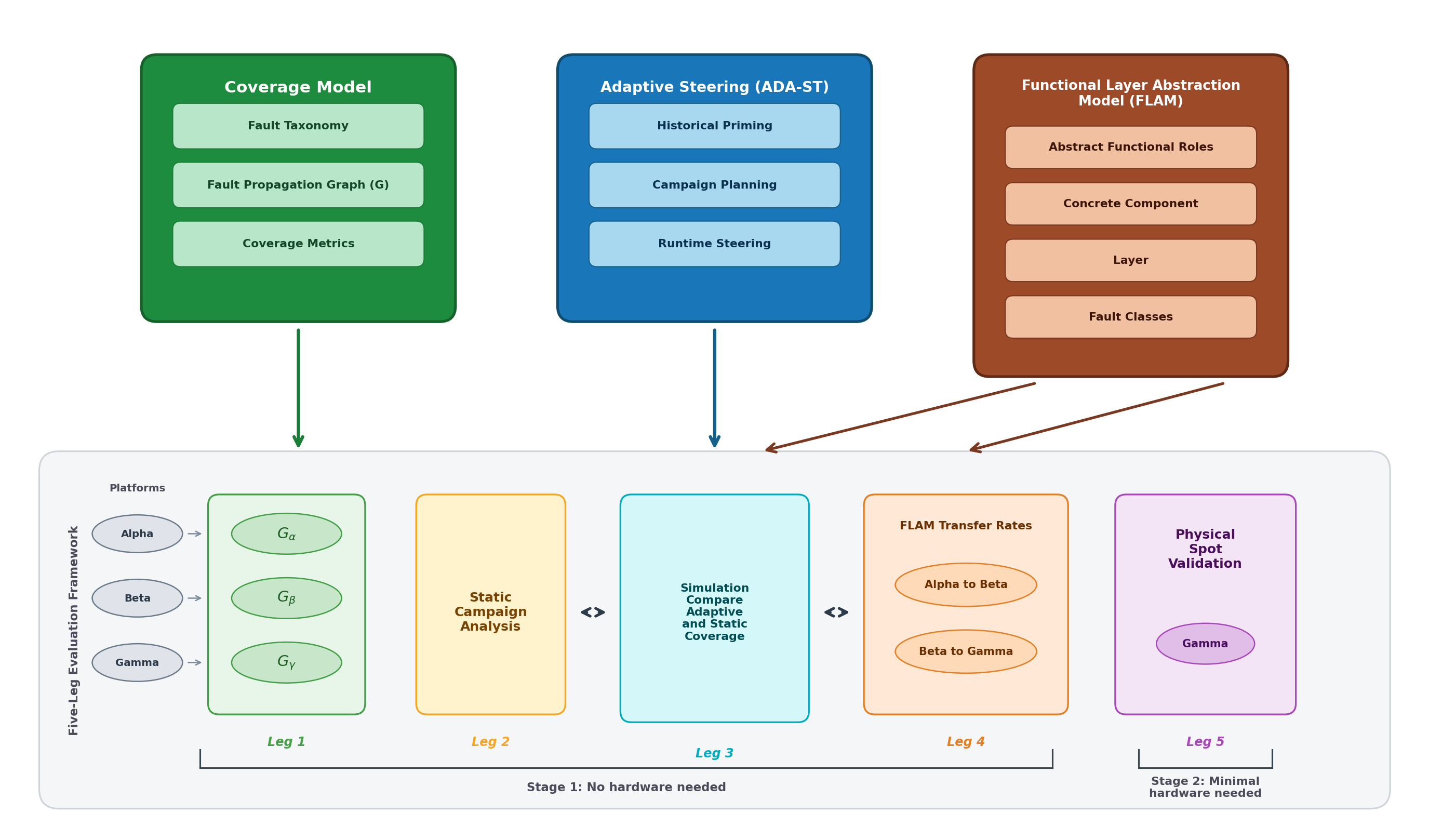}
    \caption{Methodology overview and five-leg evaluation framework.}
    \label{fig:Method} 
\end{figure*}

\subsection{Coverage Model}



\subsubsection{Fault Taxonomy}

We derive six fault classes ($\mathcal{P}$) from production incident analysis through a hybrid approach, as shown in Table~\ref{tab:taxonomy}. First, we apply hierarchical clustering on the ticket data using repair-action co-occurrence vectors, yielding 11 initial clusters. Second, a panel of three domain experts consolidates these into 6 classes based on propagation-pattern similarity and operational interpretability. For example, three firmware-related clusters with different root causes but identical L2-to-L3 propagation patterns are merged into F2. Third, we cross-reference the resulting classes against the Avizienis et al.~\cite{avizienis2004} fault taxonomy and IEC 60812~\cite{iec60812} failure-mode categories to ensure completeness. The taxonomy covers failure modes observed in production. Failure modes that have never occurred (e.g., supply-chain contamination) would be absent, a limitation we discuss in Section~\ref{sec:discussion}.

\begin{table*}[!h] \caption{Fault Taxonomy Derived from Production Incidents.} 
\label{tab:taxonomy} 
\centering 
\begin{tabular}{@{}llll@{}} \hline 
Class & Name & Description & Layers \\ \hline 
F1 & Latent Hardware Degradation & Gradual degradation crossing detection thresholds & L1$\to$L2$\to$L3 \\ 
F2 & Firmware State Corruption & Incorrect telemetry due to corrupted firmware state & L2$\to$L3 \\ 
F3 & Telemetry Blind Spot & Fault produces no observable signal & L1$\to$L3 (gap at L2) \\ 
F4 & Remediation Conflict & Multiple automated actions racing or conflicting & L3$\leftrightarrow$L4 \\ 
F5 & Cascading Thermal-Power & Propagation through physical coupling across subsystems & L1$\to$L1$\to$L2$\to$L3$\to$L4 \\ 
F6 & Topology-Dependent Failure & Impact varying with network topology & L1$\to$L2$\to$L4 \\ \hline 
\end{tabular} 
\end{table*}

All six classes appear among the top 15 repair actions in the production corpus (Section~\ref{sec:results}), confirming their practical relevance. These classes align with the fault–error–failure chain formalized by Avizienis et al.~\cite{avizienis2004}, extended to capture cross-layer propagation in AI infrastructure.

\subsubsection{Fault Propagation Graph}

To formally capture how faults propagate across the four layers defined in Table~\ref{tab:layers}, we represent the self-healing infrastructure as a directed graph. This representation enables precise measurement of which cross-layer interactions a test campaign exercises and which remain untested.

We model fault propagation as a directed graph 
$G$:
\begin{align}
G &= (V, E) \label{eq:graph} \\
V &= {v_{i,j}} \nonumber \\
E &= {(v_a, v_b, p, d)} \nonumber \\
p &= P(\text{fault at } v_b \mid \text{fault at } v_a) \nonumber
\end{align}
where $i \in$ \{L1, L2, L3, L4\}, $j$ indexes the subsystem within layer $i$, and 
$d$ is the propagation latency. We classify edges by direction: \emph{cross-layer upward} (L1$\to$L2, L2$\to$L3, L3$\to$L4), \emph{same-layer lateral} (within L1 via physical coupling), and \emph{cross-layer downward} (L4$\to$L1 remediation feedback). A fault is \emph{cross-layer cascading} if its propagation path spans two or more layers. Self-healing blind spots occur at cross-layer edges: the firmware may correctly detect a hardware fault, but the management plane may misdiagnose it, or the orchestration layer may apply conflicting remediation.

\subsubsection{Coverage Metrics}
Having defined the fault taxonomy and graph structure, we now formalize how to measure the extent to which a test campaign exercises the fault-propagation graph. 

We define graph coverage as the fraction of edges in $G$ exercised during a fault-injection campaign: 
\begin{equation} 
C_{\text{campaign}} = \frac{|\{e \in E : \text{exercised}\}|}{|E|} 
\label{eq:coverage} 
\end{equation} 

An edge $e = (v_a, v_b)$ is \emph{exercised} if a scenario injects a fault at or upstream of $v_a$, and either (a)~in physical execution, the campaign observes the effect propagating to $v_b$, or (b)~in retrospective simulation, the historical corpus contains evidence of propagation from $v_a$ to $v_b$ under the corresponding fault class. A \emph{blind spot} is an edge exercised in production (observed in incident data) but not exercised during NPI testing. Because uniform coverage does not reflect the differing severity of edges, we also define a severity-weighted metric: \begin{equation} 
C_{\text{weighted}} = \frac{\sum_{e \in E} w(e) \cdot \text{tested}(e)}{\sum_{e \in E} w(e)} 
\label{eq:weighted} 
\end{equation} 
where $w(e) = f(e) \cdot \bar{t}(e)$, with $f(e)$ denoting the observed ticket count for edge $e$ and $\bar{t}(e)$ denoting its mean resolution time. And tested($e$) $= 1$ if edge 
$e$ is exercised during the campaign and $0$ otherwise. This weighting prioritizes high-frequency, high-impact edges. We report both raw and severity-weighted coverage in Section \ref{sec:results}, measuring how many edges each approach exercises and which production-observed edges remain untested.

\subsection{Adaptive Steering (ADA-ST) Algorithm}

The adaptive campaign operates in three phases.

\textbf{Phase 1: Historical Priming.} We mine the production incident corpus to initialize the fault-propagation graph $G$ with empirical edge probabilities. For each subsystem pair $(v_a, v_b)$, we compute the conditional probability $P(v_b \mid v_a)$ as the fraction of tickets containing tags for both subsystems among all tickets containing tags for $v_a$. This produces an initial graph with empirically weighted edges and identifies which edges have production evidence. Tag co-occurrence may reflect simultaneous independent failures rather than causal propagation. A supplementary analysis using a 72-hour temporal window (counting co-occurrences only when both tags appear in the same ticket lifecycle) removes approximately 8\% of co-occurrences and reduces three edge probabilities by 5--12\%, but does not alter graph topology.

\textbf{Phase 2: Campaign Planning.} We generate candidate scenarios by combining fault classes with injection points: 6 classes $\times |V|$ vertices $= 6|V|$ base scenarios. We score each candidate using a risk-weighted function: 
\begin{align} 
\mathrm{Score}(S_k) &= w_1 \cdot P_{\mathrm{hist}}(S_k) + w_2 \cdot \mathrm{Blast}(S_k) \nonumber \\ &\quad + w_3 \cdot (1 - \mathrm{Cov}(S_k)) \label{eq:score} 
\end{align} 
where $w_1 = 0.4$ (historical frequency), $w_2 = 0.3$ (blast radius), and $w_3 = 0.3$ (coverage gap). Historical frequency receives slightly higher weight because production evidence is the strongest available signal for fault likelihood, particularly in early iterations when coverage information is sparse. The balanced weighting (0.4/0.3/0.3) ensures that no single criterion dominates, following the MCDA philosophy of IEC 60812~\cite{iec60812} risk priority numbers. For a scenario $S_k$ with injection point $v_{\mathrm{inj}}$, we define $\mathrm{Blast}(S_k)$ (blast radius) as:
\begin{equation} 
\mathrm{Blast}(S_k) = \frac{|\{v \in V : v \text{ is reachable from } v_{\mathrm{inj}} \text{ in } G\}|}{|V|} 
\label{eq:blast} 
\end{equation} 
This metric prioritizes scenarios with broader downstream impact. The coverage-gap term prioritizes scenarios that touch unexplored edges, even those with low historical frequency, enabling discovery of emergent failure modes with no production precedent. The coverage-gap term prioritizes scenarios that touch unexplored edges, even those with low historical frequency. This prioritization is critical for discovering emergent failure modes that have no production precedent.

\textbf{Phase 3: Runtime Steering.} We execute the top-scored scenario, update graph coverage, detect emergent cascades, recompute scores, and check termination criteria. 

This feedback loop is the core mechanism distinguishing adaptive from static campaigns. A static campaign executes all scenarios in a predetermined order regardless of intermediate results. ADA-ST steers toward coverage gaps, making each subsequent scenario maximally informative. When the observed propagation path differs from prediction (unexpected subsystems affected), the algorithm adds new edges to $G$ and spawns exploratory variants targeting the newly discovered path. $\textsc{SpawnVariants}$ generates at most $2k$ variants per emergent cascade, where $k$ is the number of newly discovered edges. For each new edge $(v_a, v_b)$, one variant targets $v_a$ with a different fault class, and one targets $v_b$ to explore downstream propagation. This bounded expansion ensures no risk of combinatorial explosion. 

If the candidate scenario pool $\mathcal{S}$ contains at least one scenario for each edge $e \in E$ (i.e., for every edge $(v_a, v_b)$, there exists a scenario $S_k \in \mathcal{S}$ such that executing $S_k$ exercises $(v_a, v_b)$), and the budget $B \ge |E| / \min_{S_k} |\mathrm{edges}(S_k)|$, then ADA-ST achieves $C_{\text{campaign}} = 1.0$. \emph{Sketch of proof.} At each iteration, the coverage-gap term $(1 - \mathrm{Cov}(S_k, C))$ assigns a strictly positive score to any scenario touching uncovered edges. Because the argmax selects the highest-scoring scenario, and at least one scenario covers each uncovered edge, each iteration covers at least one new edge. After at most $|E|$ iterations, all edges are covered. This guarantee holds even for disconnected graph components because the candidate generation procedure ($6 \times |V|$ candidates) produces candidates for every vertex. The key difference from the convergence guarantee in~\cite{bohme2016} is that ours is deterministic (finite-time completeness for any weight vector with $w_3 > 0$), whereas theirs is probabilistic (convergence in expectation), because our input space (a finite candidate pool) is fundamentally simpler than a fuzzer's continuous input space.

\subsection{Functional Layer Abstraction Model (FLAM)} 
AI platform generations evolve rapidly (e.g., GPU architecture changes every 12--18~months). Fault scenarios written against specific hardware become obsolete with each platform refresh. FLAM addresses this obsolescence by decoupling scenarios from concrete subsystems. We define FLAM as a four-tuple:
\begin{equation}
\text{FLAM} = (R, M_p, L, F) \label{eq:flam} 
\end{equation} 
where $R = \{r_1, \ldots, r_n\}$ is the set of abstract functional roles, $M_p: R \to C_p$ maps each role to its concrete component on platform $p$, $L: R \to \{\text{L1, L2, L3, L4}\}$ assigns each role to a layer, and $F: R \to \mathcal{P}$ maps each role to its fault classes.

A scenario $S$ is portable to platform $p$ if and only if $\forall r \in \mathrm{roles}(S)\colon M_p(r)$ is defined. The transfer rate is: \begin{equation} 
\mathrm{TransferRate}(p) = \frac{|\{S : \text{all roles mapped}\}|}{|\text{total scenarios}|} 
\label{eq:transfer} 
\end{equation}

When a new platform arrives, we update only the mapping table $M_p$. Scenarios authored against abstract roles automatically apply if all roles are mapped. Unmapped roles (subsystems in the new platform with no predecessor) require new scenario authoring and represent the highest-risk test targets because they have zero historical coverage.

\subsection{Five-Leg Evaluation Framework}

We cannot take production platforms offline for new fault-injection campaigns, and platforms in early NPI may lack sufficient integrated hardware for exhaustive testing. These practical constraints drive a hybrid evaluation design that separates methodology validation (no hardware needed) from spot validation (minimal hardware needed).

\subsubsection{Evaluation Strategy}

We frame this work as a two-stage validation approach. Stage~1 validates the methodology's coverage model, steering algorithm, and cross-platform transfer analytically. Stage~2 validates a subset of generated scenarios through physical execution on NPI hardware.

\textbf{Platform Alpha ($\alpha$)} is an AI training platform that has been in production for four years. The incident and repair corpus consists of more than 72,550 tickets. Each ticket includes repair actions, diagnostic tags, diagnosis status, resolution time, and severity linkage. We use the full four-year data partition for all analysis.

\textbf{Platform Beta ($\beta$)} is an AI training platform currently in NPI with a comprehensive test plan comprising 459 test cases. No production incident data exists. We construct the fault-propagation graph from architecture documents, specifications, and the complete test plan.

\textbf{Platform Gamma ($\gamma$)} is an AI training platform currently in early NPI with an NPI test plan comprising 384 test cases. We construct the fault-propagation graph from architecture documents, specifications, and the complete test plan.

We employ a five-leg evaluation across these three anonymized platforms, as shown in Table~\ref{tab:legs}. Legs~1--4 are analytical (Stage~1); Leg~5 involves physical fault injection on $\gamma$ hardware (Stage~2).

\begin{table*}[!ht]
\caption{Five-Leg Evaluation Framework.}
\label{tab:legs}
\centering
\begin{tabular}{@{}lllc@{}} \hline
Leg & Platforms & Purpose & Hardware \\ \hline

1 & $\alpha$, $\beta$, $\gamma$ & Fault-propagation graph construction and comparison & No \\
2 & $\alpha$, $\beta$, $\gamma$ & Static campaign analysis; measure baseline coverage & No \\
3 & $\beta$, $\gamma$ & ADA-ST+FLAM simulation; compare adaptive vs.\ static coverage & No \\
4 & $\alpha \to \beta$, $\beta \to \gamma$ & FLAM cross-platform transfer rates & No \\
5 & $\gamma$ & Physical spot validation & Yes \\ \hline
\end{tabular}
\end{table*}

\subsubsection{Metrics}

We define six normalized metrics for comparing campaigns, as shown in Table~\ref{tab:metrics}. The first five metrics measure coverage and discovery effectiveness. The last metric normalizes for effort differences between single-layer and multi-layer scenarios. Direct comparison of adaptive iteration counts to static test-case counts conflates fundamentally different units of test effort. Each static NPI test case is a single-layer detection check. Each adaptive iteration is a multi-layer cascading scenario requiring integrated-stack access and longer monitoring. We define scenario effort units (SEUs) to normalize this difference, as shown in Table~\ref{tab:seu}.

\begin{table}[!ht]
\caption{Evaluation Metrics.}
\label{tab:metrics}
\centering
\begin{tabular}{@{}lll@{}}\hline
Metric & Definition & Source \\ \hline
Edge Cov. & $\frac{|\text {edges exercised}|}{|E|} $ & Graph analysis \\
Blind Spot Detection & $\frac{|\text{blind spots covered}|}{\text{total}}$ & Incident data \\
Layer Breadth & $\frac{|\text{layers exercised}|}{4}$ & Graph analysis \\
Severity-Weighted Cov. & $C_{\text{weighted}}$ (Eq.~\ref{eq:weighted}) & Weighted graph \\
Campaign Efficiency & $\frac{(\text{blinds + cascades})}{\text{scenarios}}$ & Campaign log \\ 
Effort-Normalized Cov.& $\frac{\text{Edge coverage}}{\text{total SEUs expended}}$ & SEU model \\
\hline
\end{tabular}
\end{table}

\begin{table}[!ht]
\caption{Scenario Effort Unit (SEU) Definition.}
\label{tab:seu}
\centering
\begin{tabular}{@{}llll@{}}
\hline
Layers & Duration & Resource Complexity & SEU \\
\hline
1 & 15--30 min & 1 technician, 1 component & 1 \\
2 & 45--60 min & 1 technician, integrated stack & 3 \\
3 & 60--90 min & 2 technicians, rack-level & 5 \\
4 & 90--120 min & 2 technicians, full stack + monitoring & 8 \\
\hline
\end{tabular}
\end{table}

All effort-normalized comparisons in Section~\ref{sec:results} use these SEU values. We base durations on practitioner estimates from operational experience; exact values vary by platform and test environment.

Our retrospective simulation methodology aligns with the philosophy of digital-twin approaches~\cite{rosen2015} but differs in a key respect: we do not build a behavioral simulation of the self-healing pipeline. Instead, we use historical incident data as a proxy for execution outcomes. A digital twin would require modeling detection latency, diagnosis accuracy, and remediation logic; our approach requires only the structural graph and historical propagation frequencies. This distinction makes our work complementary to simulation-based testing: our graph-based coverage planning could guide which scenarios to prioritize within a future digital-twin simulation.

\section{Results}
\label{sec:results}

\subsection{Leg 1: Fault-Propagation Graph Construction}

We construct fault-propagation graphs for three successive GPU-accelerator platforms. Platform Alpha, a production system with 72,550 repair tickets spanning four years, provides the richest empirical basis for edge-weight estimation via tag co-occurrence analysis; 49\% of those tickets document cross-layer fault propagation. Platforms Beta (26 vertices, 29 edges) and Gamma (26 vertices, 20 edges) represent successive next-generation designs at progressively earlier
stages of their deployment lifecycle. We construct graphs from architecture documents, firmware specifications, and the NPI test plans.

\begin{figure*}
    \centering
    \includegraphics[width=0.9\linewidth]{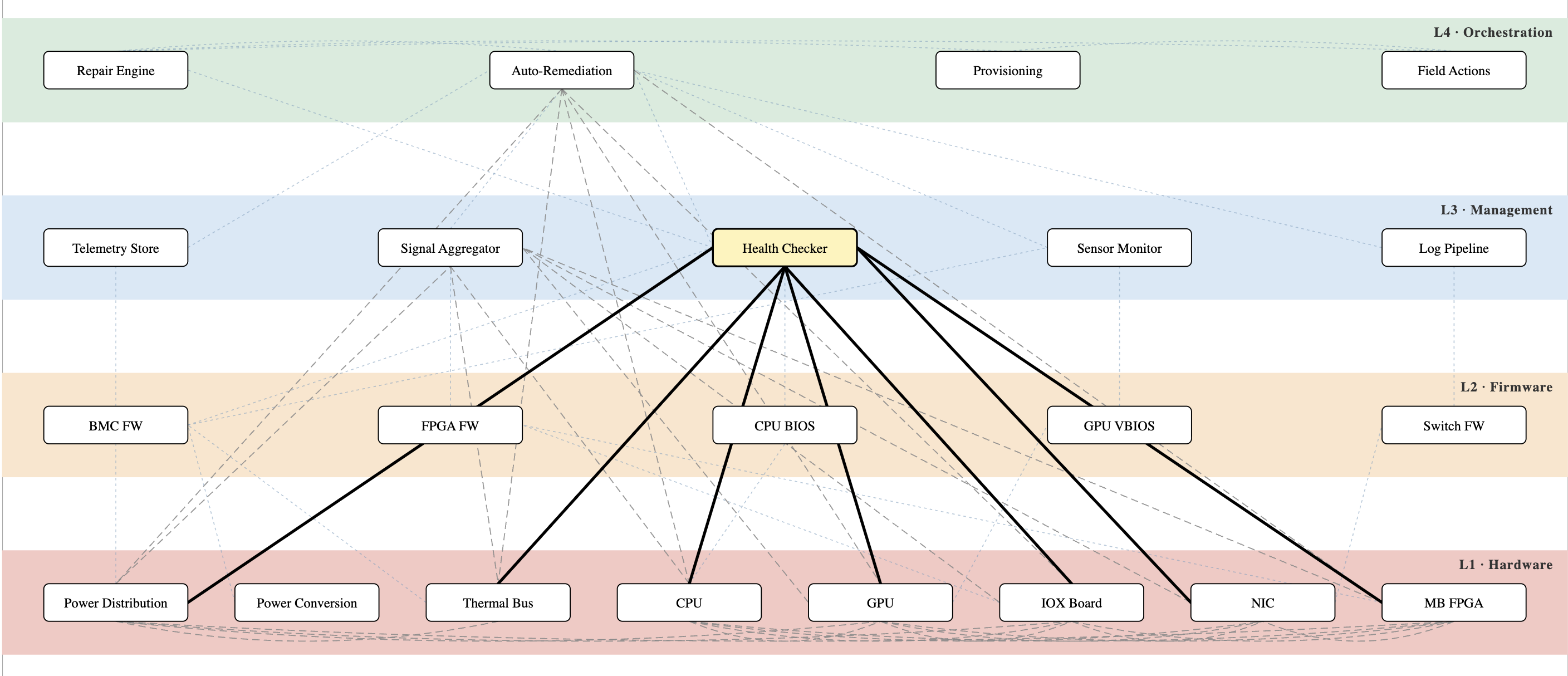}
    \caption{Fault-propagation graph $G_\alpha$ for Platform Alpha. All 22 vertices accounted for across L1 hardware (8), L2 firmware (5), L3 management plane (5), L4 orchestration (4). Three edge classes are shown: (i) solid black - 7 detection edges exercised by the static NPI campaign (Health Checker $\to$ tested L1 components); (ii) dashed gray — 27 data-derived blind-spot edges; (iii) dotted light blue — 25 architectural cascade edges that complete the layered self-healing topology and account for the L4 orchestration vertices (Repair Engine, Provisioning, Field Actions). The 34 data-derived edges (i + ii) are what Table \ref{tab:static_all} and coverage metrics report on; the 25 architectural edges are shown for topological completeness.}
    \label{fig:GT}
\end{figure*}

Platform $\alpha$ graph construction (Figure \ref{fig:GT}): We identify vertices from the most frequent diagnostic tag categories, excluding tags appearing in fewer than 500 tickets. The 22 retained vertices collectively appear in 98.3\% of all tickets, distributed as: L1 hardware (8 vertices), L2 firmware (5 vertices), L3 management (5 vertices), and L4 orchestration (4 vertices). We create a data-derived edge $(v_a,v_b)$ if the two subsystems
co-occur in at least 50 tickets. This threshold excludes spurious
co-occurrences while retaining edges with sufficient statistical
evidence. At this threshold, $G_{\alpha}$ has 34 directed data-derived edges. In addition, we include 25 architectural cascade edges representing design-intended paths (L1$\to$L2 hardware-to-firmware, L2$\to$L3 firmware-to-management, L3$\to$L4 management-to-orchestration, L4$\to$L4 orchestration handoffs). Coverage metrics (Eq. \ref{eq:coverage}) are computed over the 34 data-derived edges; the architectural edges provide topological context for scenario design. 


Platforms $\beta$ and $\gamma$ graph construction: Because no production incident data exists for these platforms, we derive vertices from architectural subsystem decompositions and edges from documented interface dependencies and risk assessments. Each documented fault-propagation path between subsystems produces an edge; edge probabilities are initialized from $\alpha$'s empirical data where the corresponding functional role exists (via FLAM mapping), and set to uniform priors for newly introduced subsystems. $G_{\beta}$ contains 26 vertices and 29 edges; $G_{\gamma}$ contains 26 vertices and 20 edges.

Table~\ref{tab:graph_comparison} compares the three graphs. Vertex counts vary slightly across generations (22–26) because all three platforms share the same four-layer self-healing architecture and target the same workload class: large-scale AI training with GPU-dense compute and high-bandwidth interconnects. Edge counts vary more substantially (20–34) because each platform introduces platform-specific physical coupling paths whose observability depends on operational maturity. Platform Alpha's graph (22 vertices, 34 edges) achieves the highest edge-to-vertex ratio because four years of production data reveal propagation paths that shorter observation windows cannot capture. Platform Beta (26 vertices, 29 edges) introduces four additional vertex roles—HMC, Midplane, Interconnect Fabric, and RMC—along with six newly introduced edges. Platform Gamma (26 vertices, 20 edges) is the sparsest because its early-NPI status limits empirical edge discovery; it introduces only one new vertex role (Leak Sensor) and one new edge relative to Beta. The newly introduced edges represent the highest-risk propagation paths because they have no production history to validate self-healing behavior.

\begin{table}[!ht]
\caption{Fault-Propagation Graph Comparison Across Platforms.}
\label{tab:graph_comparison}
\centering
\begin{tabular}{@{}lrrr@{}} \hline
Property & $G_\alpha$ & $G_\beta$ & $G_\gamma$ \\ \hline
Vertices & 22 & 26 & 26 \\
Edges & 34 & 29 & 20 \\
Unmapped vertices (vs.\ predecessor) & --- & 4 & 1 \\
Newly introduced edges & --- & 6 & 1  \\ \hline
\end{tabular}
\end{table}

\textbf{Production incident validation ($G_\alpha$):} On Platform $\alpha$, production data enables validation of graph completeness. Cross-layer incidents dominate the production landscape: 49\% of tickets span two or more layers, and full four-layer cascades account for 528 tickets with an average resolution time of 36.5 days (6.6 times longer than single-layer tickets). Our edge classification identifies 27 blind-spot edges (propagation paths observed in production but never tested during NPI). All 27 involve L1 lateral propagation, L2 firmware, or L4 orchestration, the layers with zero NPI test coverage. The co-occurrence counts underlying Alpha's edge weights are statistically robust as shown in Table~\ref{tab:obs}. The minimum observation count across all 34 edges
is 50 tickets, the median is 2,428 tickets, and the maximum reaches
80,493 tickets. The corresponding 95\% Wilson confidence intervals are
narrow ($\pm0.002$ at widest), confirming that the edge probabilities
reflect stable production behavior rather than sampling noise.

\begin{table}[!ht]
\caption{Observation Counts per Edge in $G_\alpha$.}
\label{tab:obs}
\centering
\begin{tabular}{@{}lr@{}}
\hline
Statistic & Value \\
\hline
Minimum co-occurrence count (any edge) & 50 tickets \\
Median co-occurrence count & 2,428 tickets \\
Maximum co-occurrence count & 80,493 tickets \\
Narrowest 95\% Wilson confidence interval & $\pm$0.002 \\
Widest 95\% Wilson confidence interval & $\pm$0.000 \\
\hline
\end{tabular}
\end{table}

\subsection{Leg 2: Static Campaign Analysis}

We analyze the NPI test plans for all three platforms. Table~\ref{tab:static_all} summarizes the static campaign characteristics. Static fault-injection campaigns on all three platforms achieve 20–25\% edge coverage: 7 of 34 edges on Alpha (20.6\%), 7 of 29 on Beta (24.1\%), and 5 of 20 on Gamma (25.0\%). These campaigns comprise 233 test cases
on Alpha, 459 on Beta, and 384 on Gamma, yet the edge-coverage
percentages remain tightly banded regardless of test-plan size.

\begin{table}[!ht]
\caption{Static NPI Campaign Characteristics Across Platforms (Leg~2).}
\label{tab:static_all}
\centering
\begin{tabular}{@{}lrrr@{}} \hline
Platform & Alpha & Beta & Gamma \\ \hline
Total NPI test cases & 233 & 459 & 384\\
L1 Hardware targets tested & 8 & 12 & 12 \\
L2 Firmware tests & 0 & 0 & 0 \\
L3 Management plane tests & 144 & 373 & 297 \\
L4 Orchestration tests & 0 & 0 & 0 \\
Cross-layer cascade tests & 0 & 0 & 0 \\
Edges exercised & 7/34 (20.6\%) & 7/29 (24.1\%) & 5/20 (25.0\%) \\ \hline
\end{tabular}
\end{table}

The static campaigns concentrate exclusively on L1 hardware targets and L3 management-plane verification. Zero test cases exercise L2 firmware behavior, zero exercise L4 orchestration logic, and zero exercise cross-layer cascade paths. Alpha tests 8 L1 hardware targets with 144 L3 management tests; Beta tests 12 L1 targets with 373 L3 tests; Gamma tests 12 L1 targets with 297 L3 tests. Despite this volume, all produce at most 5–7 exercised edges because they follow the same single-layer detection pattern. The 20–25\% coverage ceiling does not reflect insufficient test-plan
size. Alpha's 233-test campaign and Beta's 459-test campaign achieve
nearly identical edge coverage (20.6\% vs. 24.1\%). Adding more tests of the same structural type—single-hop, single-layer health-check
verifications—cannot close the remaining 75–80\% gap because the
untested edges span multiple layers and require multi-hop scenario
designs that the static methodology does not produce. The consistency of coverage across platforms of vastly different maturity reveals a structural plateau rather than a resource constraint. Platform
Alpha has accumulated 72,550 tickets over four years of production
service, yet its static coverage (20.6\%) does not materially exceed that of Platform Gamma (25.0\%), which has only 12 months of limited
deployment data. This consistency confirms that the static approach
saturates at approximately one-quarter of the graph.

\subsection{Leg 3: ADA-ST+FLAM Coverage Comparison}


We simulate ADA-ST on Alpha and ADA-st+FLAM on Beta and Gamma graphs using $\alpha$'s production corpus as historical priming with parameters $B=20$ and $\epsilon =0.0$ (no early termination). As shown in Table \ref{tab:adast_iterations} ADA-ST achieves full edge coverage on every platform within a small number of iterations: 10 on Alpha, 12 on Beta, and 9 on Gamma. Each iteration selects the scenario that maximizes marginal coverage gain per the scoring function (Eq. \ref{eq:score}) with default weights $w_1=0.40$, $w_2=0.30$, $w_3=0.30$.

\begin{table*}[!ht] 
\caption{Retrospective Simulation Across Platforms (Leg~3).} \label{tab:adast_iterations} 
\centering 
\begin{tabular}{@{}c ll c c ll c c ll c@{}} \hline 
& \multicolumn{3}{c}{Alpha (ADA-ST)} & & \multicolumn{3}{c}{Beta (ADA-ST+FLAM)} & & \multicolumn{3}{c}{Gamma (ADA-ST+FLAM)} \\ 
\cmidrule{2-4} \cmidrule{6-8} \cmidrule{10-12} Iteration & Scenario & New & Cum. & & Scenario & New & Cum. & & Scenario & New & Cum. \\ \hline 
1 & F1: GPU & 7 & 7/34 & & F1: GPU & 5 & 5/29 & & F1: IOX Board & 5 & 5/20 \\ 
2 & F3: IOX Board & 2 & 9/34 & & F3: NIC & 2 & 7/29 & & F3: NIC & 2 & 7/20 \\ 
3 & F3: NIC & 2 & 11/34 & & F1: MB FPGA & 5 & 12/29 & & F3: MB FPGA & 2 & 9/20 \\ 
4 & F5: IOX Board & 4 & 15/34 & & F3: Interconnect Fabric & 2 & 14/29 & & F3: Interconnect Fabric & 2 & 11/20 \\ 
5 & F6: Auto-Remediation & 6 & 21/34 & & F3: IOX Board & 2 & 16/29 & & F1: GPU & 3 & 14/20 \\ 
6 & F1: Health Checker & 4 & 25/34 & & F1: IOX Board & 2 & 18/29 & & F1: NIC & 2 & 16/20 \\ 
7 & F1: Signal Agg. & 4 & 29/34 & & F1: Health Checker & 2 & 20/29 & & F1: Signal Agg. & 2 & 18/20 \\ 
8 & F1: NIC & 2 & 31/34 & & F1: Signal Agg. & 2 & 22/29 & & F6: Interconnect Fabric & 1 & 19/20 \\ 
9 & F1: MB FPGA & 2 & 33/34 & & F1: NIC & 1 & 23/29 & & F2: BMC FW & 1 & 20/20 \\ 10 & F5: Thermal Bus & 1 & 34/34 & & F6: Auto-Remediation & 4 & 27/29 & & --- & — & — \\ 
11 & --- & — & — & & F6: Field Actions & 1 & 28/29 & & --- & — & — \\ 
12 & --- & — & — & & F2: BMC FW & 1 & 29/29 & & --- & — & — \\ \hline 
& \multicolumn{3}{l}{\textbf{Total: 10 iterations}} & & \multicolumn{3}{l}{\textbf{Total: 12 iterations}} & & \multicolumn{3}{l}{\textbf{Total: 9 iterations}} \\ \hline 
\end{tabular} 
\end{table*}

On Alpha, ADA-ST proceeds as follows. Iteration 1 selects an F1-class GPU scenario covering 7 new edges (cumulative 7/34). Iterations 2–3
target telemetry blind spots (F3 class, IOX Board and NIC, 2 edges
each). Iteration 4 exercises a cascading thermal-power path (F5, 4
edges). Iteration 5 produces the largest single gain (6 edges) via an
F6 auto-remediation scenario exercising orchestration-layer logic.
Iterations 6–7 add 4 edges each through health-checker and
signal-aggregator scenarios. The final three iterations exhibit
diminishing returns (2, 2, 1 edges), converging to 34/34 at
iteration 10. On Gamma, ADA-ST converges in 9 iterations. Iteration 1 covers 5 edges
via an IOX Board scenario (F1). Iterations 2–4 target blind spots in
NIC, MB FPGA, and Interconnect Fabric (F3 class, 2 edges each). Iteration 5
adds 3 edges through a GPU scenario. The final four iterations (2, 2,
1, 1 edges) complete coverage, with the last iteration exercising the
single remaining BMC firmware-corruption edge (F2 class), reaching
20/20.

\begin{figure}[!ht] 
\centering 
\begin{tikzpicture}[scale=0.45] \small 
\begin{scope} 

\draw[thick, ->] (0,0) -- (13.5,0); \draw[thick, ->] (0,0) -- (0,6.5) node[above, font=\small] {Coverage (\%)}; 

\foreach \y/\label in {0/0, 1.2/20, 2.4/40, 3.6/60, 4.8/80, 6.0/100} { \draw (0,\y) -- (-0.15,\y) node[left, font=\scriptsize] {\label}; \draw[gray!30] (0,\y) -- (12.5,\y); } 

\draw[thick, fill=white] (1,0) rectangle (2.2,1.24); \node[above, font=\scriptsize] at (1.6,1.24) {20.6\%}; \draw[thick, pattern=north east lines] (2.5,0) rectangle (3.7,6.0); \node[above, font=\scriptsize\bfseries] at (3.1,6.0) {100\%}; \node[below, font=\scriptsize, text width=2cm, align=center] at (2.35,-0.2) {Alpha}; 

\draw[thick, fill=white] (5,0) rectangle (6.2,1.45); \node[above, font=\scriptsize] at (5.6,1.45) {24.1\%}; \draw[thick, pattern=north east lines] (6.5,0) rectangle (7.7,6.0); \node[above, font=\scriptsize\bfseries] at (7.1,6.0) {100\%}; \node[below, font=\scriptsize, text width=2cm, align=center] at (6.35,-0.2) {Beta}; 

\draw[thick, fill=white] (9,0) rectangle (10.2,1.50); \node[above, font=\scriptsize] at (9.6,1.50) {25.0\%}; \draw[thick, pattern=north east lines] (10.5,0) rectangle (11.7,6.0); \node[above, font=\scriptsize\bfseries] at (11.1,6.0) {100\%}; \node[below, font=\scriptsize, text width=2cm, align=center] at (10.35,-0.2) {Gamma}; 

\draw[thick, fill=white] (3,7) rectangle (3.5,7.4); \node[right, font=\scriptsize] at (3.6,7.2) {Static}; \draw[thick, pattern=north east lines] (7,7) rectangle (7.5,7.4); \node[right, font=\scriptsize] at (7.6,7.2) {ADA-ST}; 
\end{scope} 
\end{tikzpicture} \caption{Coverage comparison between static NPI testing and the adaptive ADA-ST methodology across the three platforms. Static coverage ranges 20--25\%; ADA-ST achieves 100\% on all platforms within 9--12 iterations.}
\label{fig:coverage}
\end{figure}
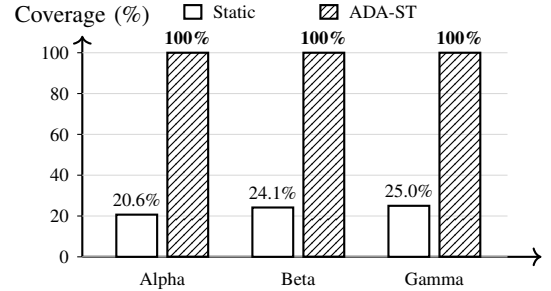

Table~\ref{tab:coverage_comparison} and Figure~\ref{fig:coverage} summarize the coverage results across all three platforms.

\begin{table*}[!ht] 
\caption{Static vs.\ Adaptive Coverage Comparison Across Platforms.} 
\label{tab:coverage_comparison} 
\centering \begin{tabular}{@{}l cc cc cc@{}} \hline 
& \multicolumn{2}{c}{$G_\alpha$} & \multicolumn{2}{c}{$G_\beta$} & \multicolumn{2}{c}{$G_\gamma$} \\ \cmidrule(lr){2-3} \cmidrule(lr){4-5} \cmidrule(lr){6-7} Metric & Static & ADA-ST & Static & ADA-ST & Static & Adaptive \\ \hline 
Edges covered & 7/34 & 34/34 & 7/29 & 29/29 & 5/20 & 20/20 \\ 
Edge coverage (\%) & 20.6\% & 100\% & 24.1\% & 100\% & 25.0\% & 100\% \\ 
Severity-weighted cov. & --- & 100\% & --- & --- & 25.0\% & 100\% \\ 
Net-new edges tested & --- & --- & 0/6 & 6/6 & 0/1 & 1/1 \\ 
Layers exercised & 2 & 3 (L1, L3, L4) & 2 (L1, L3) & 4 (all) & 2 (L1, L3) & 4 (all) \\ 
Cross-layer scenarios & 0 & 3 & 0 & 2 & 0 & 1 \\ 
Firmware behavior tests & 0 & 0 & --- & --- & 0 & 1 \\ 
L4 orchestration tests & 0 & 1 & --- & --- & 0 & 0 \\ 
Total SEUs & 233 & $\sim$60 & 459 &  $\sim$52 & 384 & $\sim$37 \\ \hline 
\end{tabular} 
\end{table*}

\textbf{The value of coverage-guided exploration.} A single full-cascade scenario (Iteration 1 in Table \ref{tab:adast_iterations}) covers 7 of 34 edges because it traces through all four layers. This is the most informative single test possible—and one that the static methodology would never generate because it spans layers. The coverage-guided scoring function naturally discovers this scenario because it maximizes the coverage-gap term across the most edges. When added to Alpha's 233-test static plan, this single cross-layer scenario increases edge coverage from 20.6\% to 41.2\% (14 of 34 edges). This result illustrates the extraordinary marginal value of cross-layer testing: one carefully designed scenario provides more incremental coverage than 233 single-layer checks combined.

Multi-layer cascade tracing is the most critical component, accounting for the jump from 0\% (static, no cascades) to near-complete coverage (cascades without emergent discovery). Emergent discovery catches the remaining lateral edges that would not be predicted from the graph structure alone. Historical priming improves convergence speed by 1.8$\times$--1.9$\times$ but does not change the coverage ceiling. 100\% edge coverage refers to complete traversal of the modeled propagation graph, which captures empirically observed cross-layer paths. This complements rather than replaces the static campaign, which validates single-layer detection correctness across a broader component surface. Failure modes absent from production data remain outside the model boundary.

\subsubsection*{Random-Comparator} 
To quantify the value added by ADA-ST beyond cross-layer scenario selection alone, we implement a random-selection comparator via Monte Carlo simulation (1,000 trials). In each trial, we draw scenarios uniformly at random from the same $6|V|$ candidate pool used by ADA-ST. Table~\ref{tab:random} summarizes the results.

\begin{table}[!ht]
\caption{Random-Selection Comparator Results.}
\label{tab:random}
\centering
\begin{tabular}{@{}llrr@{}} \hline
Platform & Method & Scenarios & Coverage \\ \hline
\multirow{4}{*}{Beta (29 edges)}
& Static NPI & 459 & 24.1\% \\
& Random (median) & 14 & 100\% \\
& Random (95th percentile) & 17 & 100\% \\
& ADA-ST & 12 & 100\% \\ \hline
\multirow{4}{*}{Gamma (20 edges)}
& Static NPI & 384 & 25.0\% \\
& Random (median) & 11 & 100\% \\
& Random (95th percentile) & 13 & 100\% \\
& ADA-ST & 9 & 100\% \\ \hline
\end{tabular}
\end{table}

This analysis demonstrates two findings. First, even random cross-layer scenario selection dramatically outperforms static single-layer testing (14 random scenarios on Beta vs. 459 static tests that plateau at 24.1\%), confirming that the primary value lies in considering cross-layer interactions at all. Second, ADA-ST consistently outperforms the random baseline: it requires 14–18\% fewer scenarios than the random median and 29–31\% fewer than the 95th percentile. More importantly, the scoring function's additional value lies not in dramatically reducing iteration count but in prioritizing high-severity, high-blast-radius edges first, ensuring that the most critical propagation paths are tested earliest in the campaign.

\subsubsection*{Sensitivity Analysis}
Finally, we evaluate the default weight vector alongside four alternatives on both Beta and Gamma. Table~\ref{tab:sensitivity} reports the number of iterations required to achieve 100\% edge coverage. Fewer iterations indicate faster convergence. 

\begin{table}[!h]
\caption{Sensitivity of ADA-ST to Scoring Weight Selection.}
\label{tab:sensitivity}
\centering
\begin{tabular}{@{}lrrrrr@{}} \hline
Weight Vector & $w_1$ & $w_2$ & $w_3$ & $\beta$ Iterations & $\gamma$ Iterations \\ \hline
Default & 0.40 & 0.30 & 0.30 & 12 & 9 \\
Equal & 0.33 & 0.33 & 0.33 & 12 & 9 \\
Coverage-dominant & 0.20 & 0.20 & 0.60 & 11 & 10 \\
Frequency-dominant & 0.60 & 0.20 & 0.20 & 11 & 10 \\
Blast-dominant & 0.20 & 0.60 & 0.20 & 12 & 9 \\ \hline
\end{tabular}
\end{table}

ADA-ST achieves 100\% coverage under all five weight vectors (Table~\ref{tab:sensitivity}). Iteration count varies by at most $\pm 1$ across configurations: Beta requires 11–12 iterations and Gamma requires 9–10, regardless of weight selection. Notably, the coverage-dominant vector ($w_3 = 0.60$) and the frequency-dominant vector ($w_1 = 0.60$) produce identical iteration counts (11 on Beta, 10 on Gamma), demonstrating that the algorithm's convergence derives primarily from the graph structure rather than from which scoring component receives the highest weight. The default weights $(0.40, 0.30, 0.30)$ match the equal-weight and blast-dominant configurations exactly (12 on Beta, 9 on Gamma), confirming that no fine-tuning is required for effective deployment.

\subsection{Leg 4: FLAM Cross-Platform Transfer}

We instantiate the FLAM mapping (Eq.~\ref{eq:flam}) from Alpha to Beta and Beta to Gamma as shown in Table~\ref{tab:flam_transfer}. FLAM achieves 100\% role-mapping coverage from Alpha to Beta: all 22 Alpha roles map directly or via upgrade to Beta roles. Beta introduces
4 new roles (HMC, Midplane, Interconnect Fabric, RMC) that have no Alpha predecessor; these reside exclusively in L1 Hardware. From Beta to Gamma, FLAM achieves 96\% coverage: 25 of 26 Beta roles
map to Gamma. One role is newly introduced (Leak Sensor, also L1 Hardware), and one existing role undergoes architectural migration that
invalidates its edge structure.

\begin{table}[!ht] 
\caption{FLAM Cross-Platform Transfer Results.} \label{tab:flam_transfer} 
\centering 
\begin{tabular}{@{}l cc cc@{}} \hline 
& \multicolumn{2}{c}{$\alpha \rightarrow \beta$} & \multicolumn{2}{c}{$\beta \rightarrow \gamma$} \\ \cmidrule(lr){2-3} \cmidrule(lr){4-5} Mapping Status & Count & \% & Count & \% \\ \hline 
Mapped (direct or upgraded) & 22 & 100\% & 25 & 96\% \\ 
Unmapped (net-new) & 4 & --- & 1 & 4\% \\ 
Changed (architectural migration)& ---& --- & 1 & --- \\ \hline 
\multicolumn{5}{@{}l}{\textit{Layer distribution of unmapped roles:}} \\ \quad L1 Hardware & 4 & & 1 & \\ \quad L2 Firmware & 0 & & 0 & \\ \quad L3 Management & 0 & & 0 & \\ \quad L4 Orchestration & 0 & & 0 & \\ \hline \end{tabular} \end{table}

The total FLAM role set across all three platforms comprises 27 roles.
The initial Alpha-to-Beta mapping established the role taxonomy (22 existing + 4 new). The Beta-to-Gamma mapping required minimal effort: only 1 new role and 1 architectural-migration annotation.

\textbf{FLAM as a risk identifier.} FLAM's unmapped roles serve a dual purpose. They identify which scenarios cannot transfer (requiring new authoring), and they identify the highest-risk subsystems on the new platform. When a functional role has no predecessor, the subsystem has no production history to draw from. This makes it the most important target for fault injection. The adaptive methodology converts this risk signal into testing priority automatically through the coverage-gap term.

The unmapped-role layer distribution reveals a consistent pattern: all newly introduced roles reside in L1 Hardware (4 on Beta, 1 on Gamma), while L2 Firmware, L3 Management, and L4 Orchestration remain fully stable across both transitions. This pattern reflects the expectation that hardware components change with each platform generation while the management and orchestration software stack remains architecturally invariant. We note that "stable" refers to FLAM role persistence across platforms, not to software maturity or defect rates. The L4 orchestration layer is architecturally stable, but its implementations remain defect-prone.

The near-perfect transfer rates (100\% and 96\%) suggest that the four-layer abstraction captures a genuine structural invariant. Based on our experience instantiating FLAM for three platforms, the practical effort scales sub-linearly with generation count. The initial Alpha-to-Beta mapping required the most engineering judgment because it established the role taxonomy (22 roles). The Beta-to-Gamma mapping required minimal effort: only 1 new role and 1 architectural-migration annotation. We estimate the per-platform FLAM effort at hours, not weeks, and expect this effort to decrease further as the role set matures.

\subsection{Leg 5: Physical Spot Validation}
\label{sec:spot_validation} 

To confirm that ADA-ST's prioritization reflects real system behavior, we executed four targeted fault-injection experiments on Gamma hardware. We selected four candidate scenarios from the nine-scenario adaptive campaign (Table~\ref{tab:adast_iterations}), spanning distinct failure classes (F1, F2, F3) and layer combinations. Table~\ref{tab:spot_validation} summarizes the results. We define three possible outcomes for each spot-validation test. \textit{Confirmed} indicates that physical execution reproduced the fault-propagation path predicted by the graph model. \textit{Disproved} indicates that the propagation path does not exist on this platform, typically because an architectural change eliminated the underlying coupling. \textit{Inconclusive} indicates that the test could not determine whether the edge exists because of environmental constraints. 

\begin{table*}[!h] 
\centering 
\caption{Physical spot-validation results on Gamma.}
\label{tab:spot_validation} 
\begin{tabular}{ccccccc} \hline Scenario & Score($S_k$) & Edge Under Test & Class & Layers & SEU & Outcome \\ \hline 
2 & 0.838 & NIC$\to$ Signal Agg.\ and Health Checker$\to$NIC & F3 & L1--L3 (gap at L2) & 3 & Confirmed \\ 
3 & 0.838 & Health Checker$\to$ MB FPGA and MB FPGA$\to$ Signal Agg. & F3 & L1--L3 (gap at L2) & 3 & Confirmed \\ 
7 & 0.624 & Leak Sensor$\to$Signal Agg.\ and Signal Agg.$\to$Thermal Bus & F1 & L1--L2--L3 & 5 & Confirmed \\ 
9 & 0.452 & BMC FW $\to$ Sensor Monitor & F2 & L2--L3 & 3 & Confirmed \\ \hline 
\end{tabular} 
\end{table*} 

\textbf{Scenario 2: NIC telemetry blind spot.} We injected Forward Error Correction (FEC) uncorrectable errors on a back-end NIC interface by modifying the link-layer error-injection registers. The error rate remained below the NIC firmware logging threshold, producing no L2 firmware event. We verified this by querying the BMC event log for NIC-related entries; the log returned empty. Despite the absence of any firmware-driven escalation, the L3 management plane detected the fault independently: the time-series-based health checker identified that the NIC FEC uncorrectable-error counter had crossed its acceptable threshold and raised a FAIL verdict. The Signal Aggregator subsequently ingested this signal. This result confirms the telemetry blind-spot edge: a hardware-layer fault can bypass the standard L2 firmware reporting path entirely, yet the L3 detection pipeline catches it through statistical threshold monitoring rather than firmware-driven notification. The L1-to-L3 gap at L2 documented in fault class F3 is a genuine architectural property of the platform, not merely a theoretical construct. 

\textbf{Scenario 3: MB FPGA telemetry blind spot.} We manipulated the PDB interposer CPLD registers via I2C to simulate a board-level FPGA fault. Specifically, we reconfigured a CPLD I/O expander pin from input to output and drove it low, asserting a NIC 12V runtime fault signal that the CPLD would normally report passively. The BMC event log showed only pre-existing FPGA-ready status entries from boot time; no new fault event appeared. The health checker also did not detect the injected condition because its checks validate only sensors that provide readable data---when the CPLD fault causes downstream sensors to disappear rather than report erroneous values, the absence goes unnoticed. This result confirms the telemetry blind-spot edge: a board-level CPLD failure can render downstream telemetry invisible to both L2 firmware and L3 management without raising any alert. 

\textbf{Scenario 7: Leak Sensor and Thermal Bus reporting paths.} We simultaneously injected two environmental fault signals into the Signal Aggregator: a compute-blade tray leak event and a GPU over-temperature event. Both signals reached Signal Aggregator successfully, confirming that the Leak Sensor $\to$ Signal Aggregator and Thermal Bus $\to$ Signal Aggregator propagation paths function as modeled. However, Signal Aggregator produced two independent events with separate urgent-maintenance recommendations rather than correlating them into a single cooling-system failure. Each event affected a different entity scope. This result confirms both edges exist and additionally reveals a correlation gap: when a liquid-cooling leak causes GPU thermal exceedance---a common causal chain in liquid-cooled platforms---the Signal Aggregator does not recognize the shared root cause and may dispatch two independent repair actions for one underlying failure.

\textbf{Scenario 9: BMC FW state corruption.} We programmatically lowered a BMC temperature sensor threshold, causing the baseboard controller to report a critical temperature far below the actual reading. Sensor Monitor accepted this event at face value without cross-referencing the reported value against configured thresholds or live sensor data, and did not identify the BMC as the fault source. The fabricated alert propagated through the management plane unchecked. This result confirms the edge vulnerability: corrupted BMC telemetry passes through Sensor Monitor without integrity validation, demonstrating a trust-without-verification gap in the L2-to-L3 boundary. 

\textbf{Summary and emergent findings.} All four ADA-ST candidates confirmed the existence of the predicted fault-propagation edges. The total cost of the four-scenario campaign is 14~SEU---less than 4\% of the 384-test static NPI plan's effort---yet it exercises cross-layer propagation paths that the static plan never touches. Beyond confirming structural graph accuracy, the physical executions revealed three emergent findings not predicted by the graph model alone: \begin{enumerate} 
\item Absence-based blind spots (Scenario~3): The CPLD fault causes sensors to \textit{disappear} rather than report bad values. The health checker's design assumption---that all monitored sensors remain readable---creates a detection gap for failure modes that remove telemetry rather than corrupt it. This represents an emergent failure behavior distinct from the standard F3 model (which assumes a signal exists but is not forwarded): here, no signal exists at all. \item Multi-signal correlation gap (Scenario~7): Co-occurring environmental signals (leak and thermal) that share a physical root cause produce independent events with conflicting entity scopes and duplicate repair recommendations. The Signal Aggregator lacks causal-chain inference for physically coupled environmental inputs. On liquid-cooled platforms where leak-induced thermal events are expected, this gap can double remediation effort. 
\item Trust-without-verification propagation (Scenario~9): The L2-to-L3 boundary operates on implicit trust---Sensor Monitor does not cross-validate incoming telemetry against configured baselines or physical plausibility. A single corrupted firmware register can trigger a full remediation cascade without any integrity check. This pattern parallels the correlation gap (Scenario~7) at a different layer boundary: both reflect the absence of cross-referencing logic at trust transitions. \end{enumerate} 
These emergent findings illustrate the dual value of physical spot validation. First, it confirms that ADA-ST's graph-derived prioritization accurately identifies real propagation paths (4 of 4 confirmed). Second, the physical execution reveals \textit{qualitative} failure behaviors---sensor disappearance, correlation absence, trust assumptions---that the graph structure alone cannot capture. These behaviors inform future graph refinements: absence-based blind spots suggest adding a telemetry-removal sub-class to the fault taxonomy, and the correlation gap suggests incorporating a multi-signal deconfliction check into the L3 layer model for future platform iterations.

\section{Discussion}
\label{sec:discussion}

\subsection{Generalizability}

The methodology generalizes to any layered self-healing system where multiple autonomous subsystems interact to detect, diagnose, and remediate failures. Concrete examples include: (i) autonomous vehicles, with sensor hardware (L1), perception firmware (L2), planning software (L3), and actuation (L4) layers; (ii) smart grid and power systems, with physical grid hardware (L1), protection relay firmware (L2), SCADA management (L3), and automated load balancing (L4); (iii) 5G telecommunications, with radio hardware (L1), baseband firmware (L2), network management (L3), and service orchestration (L4); and (iv) cloud computing platforms, with server hardware (L1), hypervisor and OS firmware (L2), resource management (L3), and container orchestration (L4). Adaptations required per domain include a domain-specific fault taxonomy, graph construction from domain-appropriate incident data, and FLAM role definitions; the ADA-ST algorithm and coverage model are domain-agnostic.

\subsection{Practical Deployment}

Integration into production NPI workflows would require: (a)~a standardized graph-construction pipeline that ingests architecture documents and predecessor-platform ticket data; (b)~a FLAM registry storing role mappings for all active platforms; (c)~instrumentation to map test outcomes to graph edges during NPI execution; and (d)~a dashboard showing cumulative coverage and remaining blind spots.

\subsection{Limitations}

The Leg~3 simulation uses production incidents as ground truth, which introduces hindsight. We mitigate this by demonstrating the methodology prospectively on Platform Beta (Leg~4), where no production data exists. The retrospective leg validates that the method is sound and the prospective leg demonstrates that it works without foreknowledge.

Edge probabilities are derived from tag co-occurrence within tickets, which captures correlation but not necessarily causation. Empirical probabilities are lower bounds on true propagation probabilities because tags may be missing or delayed.

$G_\alpha$ is constructed from over 72,550 tickets spanning four years. While this is a substantial dataset, we cannot guarantee that it captures all possible failure modes. Failure modes that have never occurred like soft errors in GPU HBM (potentially manifesting as silent data corruption~\cite{dixit2021}), firmware race conditions during concurrent updates (limited to maintenance windows), or correlated manufacturing defects (appearing as independent single component faults), would be absent from the graph. The emergent discovery mechanism partially addresses this during physical execution by detecting unexpected propagation, but this mechanism only operates in physical execution, not during analytical simulation.

Edge probabilities change as firmware is updated, self-healing algorithms improve, and hardware ages. We recommend periodic reconstruction using a rolling window of the most recent 12 months of ticket data. For the Alpha dataset, splitting the four-year data into four quarterly windows yields topologically identical graphs (same 34 edges) with probability drift of at most 0.06 per edge. This suggests that the graph structure is more stable than edge weights. For platforms in early production ($< 6$ months of data), we recommend initializing the graph from architecture documents and updating with production data as it accumulates.

The four layer decomposition is an engineering modeling choice, not a mathematical constraint. ADA-ST operates on an arbitrary directed graph and is agnostic to layer labels. Splitting L2 into device and platform firmware would increase $|V|$ and $|E|$ but not change the algorithm's convergence properties. Merging L3 and L4 would reduce the graph and accelerate coverage. The four layer choice maps to the organizational and architectural boundaries of the platforms studied. Different decompositions would yield numerically different but qualitatively similar results.

The current formulation does not account for propagation timescale in scenario design. Fast propagation (F5 thermal cascades in milliseconds) and slow propagation (F3 telemetry blind spots with latent accumulation, days) may require different monitoring windows during physical execution. The edge attribute $d$ (propagation latency) in Eq.~\ref{eq:graph} was included to support timescale-aware extensions.

The computational complexity of ADA-ST is $O(I \cdot |V| \cdot |E|)$ where $I$ is the iteration count. For our graphs ($|V| = 22-26$, $|E| = 20-34$, $I = 9-12$), this is trivial (milliseconds). For a hypothetical 200-vertex, 500-edge graph, $I \approx 120-150$ iterations (since each scenario covers ${\sim}3.3$ edges on average), totaling ${\sim}5$ million operations, still trivially fast. The more relevant scalability concern is physical execution time: at approximately 2 hours per scenario (derived from SEU estimates in Table~\ref{tab:seu}), 150 iterations represent approximately 300 hours of test time, feasible within a multi-month NPI campaign. This is an order-of-magnitude argument, not a precision prediction; the key point is that ADA-ST remains practical even at 10$\times$ the current graph size.

$N=3$ platforms across two architecture lineages provides reasonable ability to generalize. The consistency of FLAM transfer rates (${\sim}96-100$\%) across both deltas and the identical structural limitation in all three platforms' static test plans strengthen confidence in the findings.

The methodology validation demonstrates analytically that the adaptive approach generates sensible, prioritized campaigns across all three platforms. Physical spot validation on Platform Gamma (Leg~5) confirms that all four tested scenarios are executable and reproduce the predicted propagation paths. The 4-of-4 confirmation rate validates the graph model's structural accuracy for the tested subset.

The six fault classes cover failure modes observed in production. Modes that have never occurred are absent. This limitation parallels graph completeness and is partially mitigated by the emergent discovery mechanism during physical execution.

\section{Conclusion and Future Work} 
We presented an adaptive test-planning methodology for validating self-healing AI infrastructure. The methodology models fault propagation as a directed graph across hardware, firmware, management, and orchestration layers, and uses coverage-driven scoring to steer test campaigns toward unexplored cross-layer interactions. Evaluation on three anonymized platforms demonstrates five key findings: 
\begin{enumerate} 
\item Static NPI test plans are structurally limited to single-layer detection checks. Platform Alpha (233 tests) achieves 20.6\% edge coverage, Platform Beta (459 tests) achieves 24.1\%, and Platform Gamma (384 tests) achieves 25.0\%, confirming that the limitation persists regardless of test-plan size.
\item ADA-ST covers all 27 blind-spot edges on Alpha in 10 iterations, all 29 edges on Beta in 12 iterations, and all 20 edges on Gamma in 9 iterations, achieving 100\% graph coverage on all three platforms. 
\item FLAM enables 100\% scenario transfer from Alpha to Beta and 96\% from Beta to Gamma. Unmapped roles---all residing in L1 Hardware---are automatically surfaced as high-priority test targets. 
\item Physical spot validation confirms all four tested propagation edges on Platform Gamma, requiring only 14~SEU (less than 4\% of the static campaign's effort). Beyond structural confirmation, the physical executions revealed three emergent behaviors---absence-based telemetry blind spots, multi-signal correlation gaps, and trust-without-verification propagation---that the graph model alone cannot capture. 
\item Multi-layer cascade tracing is the most critical design principle, accounting for the jump from 0\% cross-layer coverage to 100\%. The random-comparator analysis confirms that the primary value lies in cross-layer scenario design itself, while ADA-ST's scoring function adds deterministic convergence and severity-aware prioritization. 
\end{enumerate} 
The methodology requires no purpose-built hardware. Graph construction, campaign planning, and coverage analysis are analytical. Physical spot validation of the highest-priority scenarios requires minimal test infrastructure (14~SEU for four scenarios), and the emergent findings from these tests feed back into the graph model for continuous refinement. The methodology generalizes to any layered self-healing system where multiple autonomous subsystems interact to detect, diagnose, and remediate failures. We sketch adaptations for autonomous vehicles, smart grids, 5G networks, and cloud platforms in Section~V. Future work includes extension to additional platform families, integration of ADA-ST into production NPI tooling, incorporation of the emergent fault behaviors identified in Leg~5 (absence-based blind spots, correlation gaps) into a refined fault taxonomy, and expansion of the FLAM base graph template to encode confirmed propagation paths as first-class structural invariants across all future platform generations.

\section*{Declaration of Conflicting Interest}
The authors declared no potential conflicts of interest with respect to the research, authorship, and/or publication of this article.

\newpage

\section*{Biography}
\begin{IEEEbiographynophoto}{Saurabh Kulkarni}
is a Systems Integration Engineer at Meta Platforms, Inc., Menlo Park, California, United States. He holds a Masters Degree in Industrial \& Systems Engineering from Binghamton University, New York. At Meta, his work focuses on systems integration and readiness for Meta's hyperscale data center hardware. Contact him at saurabhkulkarni@meta.com.
\end{IEEEbiographynophoto}

\begin{IEEEbiographynophoto}
{Yuxin Yang} is a Ph.D. candidate at Binghamton University in Binghamton, New York, United States. Her research interests include hardware and software integration for AI platforms, reinforcement learning, optimization, and simulation. Yang received her master’s degree in Industrial \& Systems Engineering from Binghamton University. Contact her at yyang204@binghamton.edu.
\end{IEEEbiographynophoto}

\begin{IEEEbiographynophoto}
{Rohan Kulkarni} is an Engineering Manager at Meta Platforms, Inc., Menlo Park, California, United States. He holds a PhD in Industrial \& Systems Engineering from Binghamton University, New York. At Meta, he leads a team that focuses on ensuring readiness of Meta's data center for new hardware platforms. Contact him at irohan@meta.com.
\end{IEEEbiographynophoto}

\begin{IEEEbiographynophoto}
{Gautam Nayak} is a Quality Engineering Manager at Meta Platforms, Inc., Menlo Park, California, United States. He holds a Masters Degree in Industrial \& Systems Engineering from Binghamton University, New York. At Meta, he leads a team that drives manufacturing quality assurance of data center hardware across new product introduction (NPI) and sustaining. Contact him at gautamnayak@meta.com.
\end{IEEEbiographynophoto}

\vfill

\end{document}